\documentclass[]{llncs}

\usepackage[dvipsnames]{xcolor}
\usepackage{microtype}
\usepackage{amsmath}
\usepackage{amssymb}
\usepackage{stmaryrd}
\usepackage{graphicx}
\usepackage{cite}
\usepackage{paralist}
\usepackage{mathpartir}
\usepackage{color}
\usepackage{hyperref}
\usepackage{float}
\usepackage{bchart}
\usepackage{pgfplots}
\usetikzlibrary{fit}
\restylefloat{figure}
\usepackage{listings}

\usepackage{listings-golang} % import this package after listings

\makeatletter
\newcommand{\xRightarrow}[2][]{\ext@arrow 0359\Rightarrowfill@{#1}{#2}}
\makeatother

\newcommand{\bi}{\begin{array}[t]{@{}l@{}}}
\newcommand{\ei}{\end{array}}
\newcommand{\ba}{\begin{array}}
\newcommand{\ea}{\end{array}}
\newcommand{\bda}{\[\ba}
\newcommand{\eda}{\ea\]}
\newcommand{\bp}{\begin{quote}\tt\begin{tabbing}}
\newcommand{\ep}{\end{tabbing}\end{quote}}

\newcommand{\mathem}{\sf}

\newcommand{\myirule}[2]{{\renewcommand{\arraystretch}{1.2}\ba{c} #1
                      \\ \hline #2 \ea}}

\newcommand{\rlabel}[1]{\mbox{(#1)}}
\newcommand{\turns}{\, \vdash \,}

\newcommand\Config[2]{(#1, #2)}
\newcommand\Override{\lhd}
\newcommand{\thread}[2]{#1 \sharp #2}
\newcommand{\clock}[2]{#1^{#2}}
\newcommand{\ploc}[2]{(#1)_{#2}}

\newcommand{\incC}[2]{{\mathem inc}(#1,#2)}
\newcommand{\maxC}[2]{{\mathem max}(#1,#2)}

\newcommand{\sep}[2]{(#1 \mid #2)}

\newcommand{\close}[1]{\mathit{close}(#1)}
\newcommand{\select}{\mathit{select}}

\newcommand{\head}{{\mathem head}}
\newcommand{\last}{{\mathem last}}
\newcommand{\pp}{\ \texttt{++}}

\newcommand{\GO}{\mbox{\mathem go}}
\newcommand{\SELECT}{\mbox{\mathem select}}

\newcommand{\SYNCMAKECHAN}{\mbox{\mathem makeChan}}
\newcommand{\SEND}[2]{#1 \leftarrow #2}
\newcommand{\RCV}[1]{\leftarrow #1}

\newcommand{\SyncChan}{\mathit{Chan}}

\newcommand{\tid}{\mbox{\mathem{tid}}}
\newcommand{\newTID}{\mbox{\mathem{tidB}}}
\newcommand{\isBuffered}[1]{\mathem{isBuffered}(#1)}

\newcommand{\pre}[1]{\mathit{pre}(#1)}
\newcommand{\post}[1]{\mathit{post}(#1)}
\newcommand{\postAsync}[2]{\mathit{postB}(#1,#2)}

\newcommand{\hash}[1]{{\mathem hash}(#1)}
\newcommand{\instr}[2]{\mathit{instr}(#1) = #2} %%{\turns #1 \leadsto #2}
\newcommand{\instrt}[1]{\mathit{instr}(#1)}
\newcommand{\retrieve}[1]{\mathit{retr}(#1)}

\newcommand{\semB}[4]{(#1, #2) \turns #3 \Downarrow #4}
\newcommand{\semP}[3]{#1 \xRightarrow{#2} #3}

\newcommand{\replay}[3]{\semP{#1}{#2}{#3}}
\newcommand{\postProc}[2]{#1 \Rightarrow #2}

\newcommand{\assign}{:=}

\newcommand{\sndEvt}[2]{#1 \sharp #2!}
\newcommand{\rcvEvt}[3]{#1 \leftarrow #2 \sharp #3?}

\newcommand{\snd}[1]{#1 !}
\newcommand{\rcv}[1]{#1 ?}

\makeatletter
\newdimen\legendxshift
\newdimen\legendyshift
\newcount\legendlines
\newcommand{\bclldist}{1mm}
\newcommand{\bclegend}[3][10mm]{%
	\legendxshift=0pt\relax
	\legendyshift=0pt\relax
	\xdef\legendnodes{}%
	\foreach \lcolor/\ltext [count=\ll from 1] in {#3}%
	{\global\legendlines\ll\pgftext{\setbox0\hbox{\bcfontstyle\ltext}\ifdim\wd0>\legendxshift\global\legendxshift\wd0\fi}}%
	\@tempdima#1\@tempdima0.5\@tempdima
	\pgftext{\bcfontstyle\global\legendxshift\dimexpr\bcwidth-\legendxshift-\bclldist-\@tempdima-0.72em}
	\legendyshift\dimexpr5mm+#2\relax
	\legendyshift\legendlines\legendyshift
	\global\legendyshift\dimexpr\bcpos-2.5mm+\bclldist+\legendyshift
	\begin{scope}[shift={(\legendxshift,\legendyshift)}]
		\coordinate (lp) at (0,0);
		\foreach \lcolor/\ltext [count=\ll from 1] in {#3}%
		{
			\node[anchor=north, minimum width=#1, minimum height=5mm,fill=\lcolor] (lb\ll) at (lp) {};
			\node[anchor=west] (l\ll) at (lb\ll.east) {\bcfontstyle\ltext};
			\coordinate (lp) at ($(lp)-(0,5mm+#2)$);
			\xdef\legendnodes{\legendnodes (lb\ll)(l\ll)}
		}
		\node[draw, inner sep=\bclldist,fit=\legendnodes] (frame) {};
	\end{scope}
}
\makeatother

\title{Trace-Based Run-Time Analysis of Message-Passing Go Programs}

\author{Martin Sulzmann and Kai Stadtm{\"u}ller}
\institute{
  Faculty of Computer Science and Business Information Systems \\
  Karlsruhe University of Applied Sciences \\
  Moltkestrasse 30, 76133 Karlsruhe, Germany\\
  \email{martin.sulzmann@hs-karlsruhe.de} \\
    \email{kai.stadtmueller@live.de} 
}

\begin{document}

\maketitle

\begin{abstract}
  We consider the task of analyzing message-passing programs
  by observing their run-time behavior.
  We introduce a purely library-based instrumentation method to trace communication events
  during execution. A model of the dependencies among events can
  be constructed to identify potential bugs.
  Compared to the vector clock method, our approach is much simpler and
  has in general a significant lower run-time overhead.
  A further advantage is that we also trace events
  that could not commit. Thus, we can infer
  more alternative communications.
  This provides the user with additional information to identify potential bugs.
  We have fully implemented our approach in the Go programming language
  and provide a number of examples to substantiate our claims.
\end{abstract}

%--------------------------------------------------------
%--------------------------------------------------------
\section{Introduction}

We consider run-time analysis of programs that employ message-passing.
Specifically, we consider the Go programming language~\cite{golang}
which integrates message-passing in the style of
Communicating Sequential Processes (CSP)~\cite{Hoare:1978:CSP:359576.359585}
into a C style language.
We assume the program is instrumented to trace communication events that took
place during program execution.
Our objective is to analyze program traces to assist
the user in identifying potential concurrency bugs.

%--------------------------------------------------------
\paragraph{Motivating Example}

\begin{lstlisting}[float={tp},captionpos={b},caption={Message passing in Go},label={lst:message-passing-go}]
 func reuters(ch chan string) { ch <- "REUTERS" } // r!
 func bloomberg(ch chan string) { ch <- "BLOOMBERG" } // b!

func newsReader(rCh chan string, bCh chan string) {
  ch := make(chan string)
  go func() { ch <- (<-rCh) }()        // r?; ch!
  go func() { ch <- (<-bCh) }()        // b?; ch!
  x := <-ch                            // ch?
}

func main() {
  reutersCh := make(chan string)
  bloombergCh := make(chan string)
  go reuters(reutersCh)
  go bloomberg(bloombergCh)
  go newsReader(reutersCh, bloombergCh) // N1
  newsReader(reutersCh, bloombergCh)    // N2
}
\end{lstlisting}

In Listing~\ref{lst:message-passing-go} we find
a Go program implementing a system of newsreaders.
The \lstinline{main} function creates two synchronous channels,
one for each news agency.
Go supports (a limited form of) type inference and therefore no type annotations are required.
Next, we create one thread per news agency via the keyword \GO.
Each news agency transmits news over its own channel.
In Go, we write \texttt{ch <- "REUTERS"} to send
value \texttt{"REUTERS"} via channel \texttt{ch}.
We write \texttt{<-ch} to receive a value via channel \texttt{ch}.
As we assume synchronous channels, both operations block
and only unblock once a sender finds a matching receiver.
We find two newsreader instances.
Each newsreader creates two helper threads that wait
for news to arrive and transfer any news that has arrived
to a common channel. The intention is that the newsreader
wishes to receive \emph{any} news whether it be from
Reuters or Bloomberg.
However, there is a subtle bug (to be explained shortly).

%--------------------------------------------------------
\paragraph{Trace-Based Run-Time Verification}

%% To catch bugs in programs we apply run-time verification (testing).
We only consider finite program runs and therefore each of the news agencies supplies
only a finite number of news (exactly one in our case) and then terminates.
During program execution, we trace communication events, e.g.~send and receive, that took place.
Due to concurrency, a bug may not manifest itself because
a certain `bad' schedule is rarely taken in practice.

Here is a possible trace resulting from a `good' program run.
\begin{verbatim}
r!; N1.r?; N1.ch!; N1.ch?; b!; N2.b?; N2.ch!; N2.ch?
\end{verbatim}
We write \texttt{r!} to denote that a send event via the Reuters channel took place.
As there are two instances of the \lstinline{newsReader} function,
we write \texttt{N1.r?} to denote that a receive event via the local channel took place
in case of the first \lstinline{newsReader} call.
From the trace we can conclude that the Reuters news was consumed
by the first newsreader and the Bloomberg news by the second newsreader.

Here is a trace resulting from a bad program run.
\begin{verbatim}
r!; b!; N1.r?; N1.b?; N1.ch!; N1.ch?; DEADLOCK
\end{verbatim}
The helper thread of the first newsreader receives the Reuters
\emph{and} the Bloomberg news.
However, only one of these messages will actually be read (consumed).
This is the bug!
Hence, the second newsreader gets stuck and we encounter a deadlock.
The issue is that such a bad program run may rarely show up.
So, the question is how can we assist the user based on the trace information resulting
from a good program run? How can we infer that alternative schedules and communications
may exist?

%--------------------------------------------------------
\paragraph{Event Order via Vector Clock Method}

A well-established approach is to derive a partial order among events.
This is usually achieved via a vector of (logical) clocks.
The vector clock method was independently developed by
Fidge~\cite{fidge1987timestamps} and Mattern~\cite{Mattern89virtualtime}.
For the above good program run, we obtain the following partial order among events.
\begin{verbatim}
r! < N1.r?         b! < N2.b?
N1.r? < N1.ch!     N2.b? < N2.ch!  (1)
N1.ch! < N1.ch?    N2.ch! < N2.ch? (2)
\end{verbatim}
For example, (1) arises because \texttt{N2.ch!} happens (sequentially) after \texttt{N2.b?}
For synchronous send/receive, we assume that receive happens after send. See (2).
Based on the partial order, we can conclude that alternative schedules are possible.
For example, \texttt{b!} could take place before \texttt{r!}.
However, it is not clear how to infer alternative communications.
Recall that the issue is that one of the newsreaders may consume both news messages.
Our proposed method is able to clearly identify this issue and
has the advantage to require a much simpler instrumentation
We discuss these points shortly. First, we take a closer look
at the details of instrumentation for the vector clock method.

Vector clocks are a refinement of Lamport's time stamps~\cite{lamport1978time}.
Each thread maintains a vector of (logical) clocks of all participating partner threads.
For each communication step, we advance and synchronize clocks.
In pseudo code, the vector clock instrumentation for event \texttt{sndR}.
\begin{lstlisting}
  vc[reutersThread]++
  ch <- ("REUTERS", vc, vcCh)
  vc' := max(vc, <-vcCh)
\end{lstlisting}
We assume that \texttt{vc} holds the vector clock.
The clock of the Reuters thread is incremented.
Besides the original value, we transmit the sender's vector clock
and a helper channel \texttt{vcCh}. For convenience, we use tuple notation.
The sender's vector clock is updated by building
the maximum among all entries of its own vector clock
and the vector clock of the receiving party.
The same vector clock update is carried out on the receiver side.

\paragraph{Our Method}

We propose a much simpler instrumentation and tracing method
to obtain a partial order among events.
Instead of a vector clock, each thread traces the events that might happen
and have happened. We refer to them as pre and post events.
In pseudo code, our instrumentation for \texttt{sndR} looks like follows.
\begin{lstlisting}
  pre(hash(ch), "!")
  ch <- ("REUTERS", threadId)
  post(hash(ch), "!")
\end{lstlisting}
The bang symbol (`!') indicates a send operation. Function \texttt{hash}
builds a hash index of channel names.
The sender transmits its thread id number to the receiver.
This is the only intra-thread overhead. No extra communication link is necessary.

Here are the traces for individual threads resulting from the above good program run.
\begin{verbatim}
R:           pre(r!); post(r!)
N1_helper1:  pre(r?); post(R#r?); pre(ch1!); post(ch1!)
N1_helper2:  pre(b?)
N1:          pre(ch1?); post(N1_helper1#ch1?)
B:           pre(b!); post(b!)
N2_helper1:  pre(r?)
N2_helper2:  pre(b?); post(B#b?); pre(ch2!); post(ch2!)
N2:          pre(ch2?); post(N2_helper2#ch2?)
\end{verbatim}
We write \texttt{pre(r!)} to indicate that a send via the Reuters channel might happen.
We write \verb+post(R#r?)+ to indicate that a receive has happened via thread \texttt{R}.
The partial order among events is obtained by a simple post-processing phase
where we linearly scan through traces.
For example, within a trace there is a strict order
and therefore
\begin{verbatim}
N2_helper2:  pre(b?); post(B#b?); pre(ch2!); post(ch2!)
\end{verbatim}
implies \texttt{N2.b? < N2.ch!}.
Across threads we check for matching pre/post events.
Hence,
\begin{verbatim}
R:           pre(r!); post(r!)
N1_helper1:  pre(r?); post(R#r?); ...
\end{verbatim}
implies \texttt{r! < N1.r?}.
So, we obtain the same (partial order) information as the vector clock approach
but with less overhead.

The reduction in terms of tracing overhead compared to the vector clock
method is rather drastic assuming a library-based tracing scheme
with no access to the Go run-time system.
For each communication event we must exchange
vector clocks, i.e.~$n$ additional (time stamp) values need to be transmitted
where $n$ is the number of threads.
Besides extra data to be transmitted, we also require an extra communication link because
the sender requires the receivers vector clock.
In contrast, our method incurs a constant tracing overhead.
Each sender transmits in addition its thread id.
No extra communication link is necessary.
This results in \emph{much} less run-time overhead as we will see later.

The vector clock tracing method can be improved assuming we extend the Go run-time system.
For example, by maintaining a per-thread vector clock
and having the run-time system carrying out the exchange of vector clocks for each
send/receive communication.
There is still the $O(n)$ space overhead.
Our method does not require any extension of the Go run-time system
to be efficient and therefore is also applicable to other languages that offer
similar features as found in Go.

A further advantage of our method is that
we also trace (via pre) events that could not commit (post is missing).
Thus, we can easily infer alternative communications.
For example, for 
\texttt{R: pre(r!); ...} there is the alternative
match \verb+N2_helper1:  pre(r?)+.
Hence, instead of \texttt{r! < N1.r?} also \texttt{r! < N2.r?} is possible.
This indicates that one newsreader may consume both news message.
The vector clock method, only traces events that could commit, post events in
our notation. Hence, the above alternative communication could not be derived.

%--------------------------------------------------------
\paragraph{Contributions}

Compared to earlier works based on the vector clock method,
we propose a much more light-weight and more informative instrumentation
and tracing scheme.
Specifically, we make the following contributions:
\begin{itemize}
\item We give a precise account of our run-time tracing method (Section~\ref{sec:instrumentation})
  for message-passing as found in the Go programming language (Section~\ref{sec:message-passing-go})
  where for space reasons we only formalize the case of synchronous channels
  and selective communications.
  
\item A simple analysis of the resulting traces allows us to detect
      alternative schedules and communications (Section~\ref{sec:trace-analysis}).
      For efficiency reasons, we employ a directed dependency graph to represent
      happens-before relations (Section~\ref{sec:dep-graph}).

 \item We show that vector clocks can be easily recovered 
       based on our tracing method (Section~\ref{sec:comparison}).
       We also discuss the pros and cons of both methods for analysis purposes.      
      
 \item Our tracing method can be implemented efficiently as a library.
              We have fully implemented the approach supporting
      all Go language features dealing with message-passing
      such as buffered channels, select with default or timeout
      and closing of channels
      (Section~\ref{sec:implementation}).
      
\item We provide experimental results measuring
      the often significantly lower overhead of our method
      compared to the vector clock method assuming based methods are implemented
      as libraries (Section~\ref{sec:experiments}).

%% MS: omit, not really      
%%      We also discuss examples of how to detect potential bugs
%%      based on our trace analysis method.
\end{itemize}
 
The online version of this paper contains
an appendix with further details.\footnote{\url{https://arxiv.org/abs/1709.01588}}

%--------------------------------------------------------
%--------------------------------------------------------
\section{Message-Passing Go}
\label{sec:message-passing-go}

\paragraph{Syntax}

For brevity, we consider a much simplified fragment of the Go programming language.
We only cover straight-line code, i.e.~omitting procedures, if-then-else etc.
This is not an onerous restriction as we only consider finite program runs.
Hence, any (finite) program run can be represented as a program consisting of straight-line code only.

\begin{definition}[Program Syntax]
  \bda{lcll}
  x,y, &\dots & &
  \mbox{Variables, Channel Names}
  \\ i,j, &\dots && \mbox{Integers}
  \\
  b & ::= &  x \mid i \mid \hash{x} \mid \head(b) \mid \last(b) \mid \textit{bs} \mid \tid   & \mbox{Expressions}
  \\ \textit{bs} & ::= & [] \mid b : \textit{bs}
  \\
  e,f & ::= & \SEND{x}{b} \mid y \assign \RCV{x}  & \mbox{Transmit/Receive}
  \\
  c & ::= & y \assign b \mid y \assign \SYNCMAKECHAN \mid \GO\ p \mid \SELECT\ [e_i \Rightarrow p_i]_{i\in I} & \mbox{Commands}
  \\
  p,q,r & ::= & [] \mid c : p   & \mbox{Program}
  \eda
\end{definition}

For our purposes,
values are integers or lists (slices in Go terminology).
For lists we follow Haskell style notation
and write $b : bs$ to refer to a list with head element $b$ and tail $bs$.
We can access the head and last element in a list via primitives $\head$ and $\last$.
We often write $[b_1,\dots,b_n]$ as a shorthand $b_1:\dots : []$.
Primitive $\tid$ yields the thread id number of the current thread.
We assume that the main thread always has thread id number $1$
and new thread id numbers are generated in increasing order.
Primitive $\hash{}$ yields a unique hash index for each variable name.
Both primitives show up in our instrumentation.

A program is a sequence of commands where commands are stored in a list.
Primitive $\SYNCMAKECHAN$ creates a new synchronous channel.
Primitive $\GO$ creates a new go routine (thread).
For send and receive over a channel we follow Go notation.
We assume that a receive is always tied to an assignment.
For assignment we use symbol $:=$ to avoid confusion with the mathematical
equality symbol $=$.
In Go, symbol $:=$ declares a new variable with some initial value.
We also use $:=$ to overwrite the value of existing variables.
As a message passing command we only support selective communication
via \SELECT.
Thus, we can fix the bug in our newsreader example.
\begin{lstlisting}
func newsReaderFixed(rCh chan string, bCh chan string) {
  ch := make(chan string)
  select {
    case x := <-rCh:
    case x := <-bCh:
  }
}
\end{lstlisting}
The \SELECT\ statement guarantees that at most one news message will be consumed
and blocks if no news are available.
In our simplified language,
we assume that the $\SEND{x}{b}$ command is a shorthand
for $\SELECT\ [\SEND{x}{b} \Rightarrow []]$.
For space reasons, we omit buffered channels, select
paired with a default/timeout case and closing of channels.
All three features are fully supported by our implementation.

\paragraph{Trace-Based Semantics}

The semantics of programs is defined via a small-step operational semantics.
The semantics keeps track of the trace of channel-based communications that took place.
This allows us to relate the traces obtained by our instrumentation
with the actual run-time traces.

We support multi-threading via a reduction
relation
$$
\semP{\Config{S}{[\thread{i_1}{p_1},\dots,\thread{i_n}{p_n}]}}{T}{\Config{S'}{[\thread{j_1}{q_1},\dots,\thread{j_n}{q_n}]}}.
$$
We write $\thread{i}{p}$ to denote a program $p$ that runs in its own
thread with thread id~$i$. We use lists to store the set of program threads.
The state of program variables,
before and after execution, is recorded in $S$ and $S'$.
We assume that threads share the same state.
Program trace $T$ records the sequence of communications that took place
during execution.
We write $\snd{x}$ to denote a send operation on channel $x$
and $\rcv{x}$ to denote a receiver operation on channel $x$.
The semantics of expressions is defined in terms a big-step semantics.
We employ a reduction relation $\semB{i}{S}{b}{v}$
where $S$ is the current state, $b$ the expression
and $v$ the result of evaluating $b$.
The formal details follow.

\begin{definition}[State] 
  \bda{lcll}
  v & ::= & x \mid i \mid [] \mid \textit{vs} & \mbox{Values}
  \\
  \textit{vs} & ::= & [] \mid v : \textit{vs}
   \\ s & ::= & v \mid \SyncChan & \mbox{Storables}
   \\
   S & ::= & () \mid (x \mapsto s) \mid S \Override S                    & \mbox{State}
  \eda
 \end{definition}
A state $S$ is either empty, a mapping, or an
  override of two states.
Each state maps variables to storables.
A storable is either a plain value or a channel.
Variable names may appear as values. In an actual implementation,
we would identify the variable name by a unique hash index.
We assume that mappings in the right operand of the map override operator $\Override$ take
precedence. They overwrite any mappings in the left operand.
That is, $(x \mapsto v_1) \Override (x \mapsto v_2) = (x \mapsto v_2)$.

\begin{definition}[Expression Semantics $\semB{i}{S}{b}{v}$]
  \bda{c}
  \myirule{S(x) = v}
          {\semB{i}{S}{x}{v}}
  \ \ \ \
  \semB{i}{S}{j}{j}
  \ \ \ \
  \semB{i}{S}{[]}{[]}
  \ \ \ \
  \myirule{\semB{i}{S}{b}{v}
           \ \semB{i}{S}{\textit{bs}}{\textit{vs}}
          }
          {\semB{i}{S}{b:\textit{bs}}{v:\textit{vs}}}
  \\
  \myirule{\semB{i}{S}{b}{v : \textit{vs}}}
          {\semB{i}{S}{\head(b)}{v}}
  \ \ \ \
  \myirule{\semB{i}{S}{b}{[v_1,\dots,v_n]}}
          {\semB{i}{S}{\last(b)}{v_n}}
  \ \ \ \ 
  \semB{i}{S}{\tid}{i}
  \ \ \ \ 
  \semB{i}{S}{\hash{x}}{x}
  \eda
\end{definition}

\begin{definition}[Program Execution $\semP{\Config{S}{P}}{T}{\Config{S'}{Q}}$]
  \bda{lcll}
  \thread{i}{p} &&& \mbox{Single program thread}
  \\ P,Q & ::= & [] \mid \thread{i}{p} : P & \text{Program threads}
  \\
  t & \assign & \sndEvt{i}{x} \mid \rcvEvt{i}{j}{x} & \text{Send and receive event}
  \\
   T & ::= & [] \mid t : T & \mbox{Trace}
  \eda

\end{definition}

We write $\semP{\Config{S}{P}}{}{\Config{S'}{Q}}$
as a shorthand for
$\semP{\Config{S}{P}}{[]}{\Config{S'}{Q}}$.

\begin{definition}[Single Step]
  
  \bda{c}
  \rlabel{Terminate} \
  \semP{\Config{S}{\thread{i}{[]} : P}}
       {}
       {\Config{S}{P}}
  \\
  \\     
  \rlabel{Assign} \
  \myirule{\semB{i}{S}{b}{v} \ \ \ S' = S \Override (y \mapsto v)}
      {
       \semP{\Config{S}{\thread{i}{(y \assign b:p)}:P}}
       {}
       {\Config{S'}{\thread{i}{p}:P}}
      }
  \\
  \\
  \rlabel{MakeChan} \
  \myirule{S' = S \Override (y \mapsto \SyncChan }
      {
       \semP{\Config{S}{\thread{i}{(y \assign \SYNCMAKECHAN:p)}:P}}
       {}
       {\Config{S'}{\thread{i}{p}:P}}
      }
  \eda
\end{definition}

\begin{definition}[Multi-Threading and Synchronous Message-Passing]
\bda{c}
  \rlabel{Go} \
  \myirule{i \not \in \{i_1,\dots,i_n\} }
        {
       \semP{\Config{S}{\thread{i_1}{(\GO\ p:p_1)}:P}}
       {}
       {\Config{S}{\thread{i}{p}:\thread{i_1}{p_1}:P}}
        }
  \\
  \\
  \rlabel{Sync} \
  \myirule{\exists l \in J, m \in K. e_l = \SEND{x}{b} \ \ \ f_m = y \assign \RCV{x}
           \ \ \ S(x) = \SyncChan
          \\ \semB{i_1}{S}{b}{v} \ \ \ S' = S \Override (y \mapsto v)
          }
          {\semP{\Config{S}{\thread{i_1}{(\SELECT\ [e_j \Rightarrow q_j]_{j\in J}:p_1)}:
                           \thread{i_2}{(\SELECT\ [f_k \Rightarrow r_k]_{k\in K}:p_2)}:P}\\}
                {[\sndEvt{i_1}{x}, \rcvEvt{i_2}{i_1}{x}]}
                {\\ \Config{S'}{\thread{i_1}{(q_l \pp\ p_1)}:\thread{i_2}{(r_m \pp\ p_2)}:P}}
          }
  \eda
\end{definition}

\begin{definition}[Scheduling]
\bda{c}
  \rlabel{Schedule} \
  \myirule{\mbox{$\pi$ permutation on $\{1,\dots,n\}$}}
          {\semP{\Config{S}{[\thread{i_1}{p_1}, \dots, \thread{i_n}{p_n}]}}
                {}
                {\Config{S}{[\thread{\pi(i_1)}{p_{\pi(1)}}, \dots, \thread{\pi(i_n)}{p_{\pi(n)}}]}}
          }
  \\
  \\
  \rlabel{Closure} \
  \myirule{\semP{\Config{S}{P}}{T}{\Config{S'}{P'}} \ \ \ \semP{\Config{S'}{P'}}{T'}{\Config{S''}{P''}}}
          {\semP{\Config{S}{P}}{T \pp\ T'}{\Config{S''}{P''}}}
  \eda
\end{definition}  

%--------------------------------------------------------
%--------------------------------------------------------
\section{Instrumentation and Run-Time Tracing}
\label{sec:instrumentation}

%% MS: omit, saving space
%% We discuss the details of instrumentation and tracing.
%% The analysis phase is considered in the following section.
For each message passing primitive (send/receive) we log two events.
In case of send,
(1) a \emph{pre} event to indicate the message is about to be sent,
and (2) a \emph{post} event to indicate the message has been sent.
The treatment is analogous for receive.
In our instrumentation, we write $\snd{x}$ to denote a single send event
and $\rcv{x}$ to denote a single receive event.
These notations are shorthands and can be expressed
in terms of the language described so far.
We use $\equiv$ to define short-forms and their encodings.
We define
$\snd{x} \equiv [\hash{x},1]$ and
$\rcv{x} \equiv [\hash{x},0]$.
That is, send is represented by the number~$1$ and receive by the number~$0$.

As we support non-deterministic selection,
we employ a list of pre events to indicate
that one of several events may be chosen
For example, $\pre{[\snd{x}, \rcv{y}]}$
indicates that there is the choice
among sending over channel $x$ and receiving over channel $y$.
This is again a shorthand notation
where we assume $\pre{[b_1,\dots,b_n]} \equiv [0,b_1,\dots,b_n]$.

A post event is always singleton as at most one of the possible
communications is chosen.
As we also trace communication partners, we assume
that the sending party transmits its identity, the thread id,
to the receiving party.
We write $\post{\thread{i}{\rcv{x}}}$
to denote reception via channel $x$ 
where the sender has thread id $i$.
In case of a post send event, we simply write
$\post{\snd{x}}$.
The above are yet again shorthands
where $\thread{i}{\rcv{x}} \equiv [\hash{x},0,i]$ and
      $\post{b} \equiv [1,b]$.

Pre and post events are written in a fresh thread local variable,
denoted by $x_{tid}$ where $tid$ refers to the thread's id number.
At the start of the thread
the variable is initialized by $x_{tid} \assign []$.
Instrumentation ensures that pre and post events are appropriately
logged. As we keep track of communication partners,
we must also inject and project messages with additional information
(the sender's thread id).

We consider instrumentation of
$
\SELECT\ [\SEND{x}{1} \Rightarrow [], y \assign \RCV{x} \Rightarrow [ \SEND{z}{y}]].
$
We assume the above program text is part of a thread with id number $1$.
We non-deterministically choose
between a send an receive operation.
In case of receive, the received value is further transmitted.
Instrumentation yields the following.

\bda{l}
[x_1 \assign x_1 \pp\ \pre{[\snd{x}, \rcv{x}]},
\\
\ \SELECT\ [\SEND{x}{[\tid, 1]} \Rightarrow [x_1 \assign x_1 \pp\ \post{\snd{x}}],
\\ \ \ \ \ \ \ \ \ \ \
y' \assign \RCV{x} \Rightarrow
  [ x_1 \assign x_1 \pp\ \post{\thread{\head(y')}{\rcv{x}}}, y \assign \last(y'),
\\ \ \ \ \ \ \ \ \ \ \ \ \ \ \ \ \ \ \ \ \ \ \ \ \ \ \ \   \SEND{z}{[\tid, y]}]]
\eda
We first store the pre events, either a read or send via channel $x$.
The send is instrumented by additionally transmitting the senders thread id.
The post event for this case simply logs that a send took place.
Instrumentation of receive is slightly more involved.
As senders supply their thread id, we introduce a fresh variable $y'$.
Via $\head(y')$ we extract the senders thread id to properly
record the communication partner in the post event.
The actual value transmitted is accessed via $\last(y')$.

\begin{definition}[Instrumentation of Programs]
  \label{def:instrumentation}    
We write $\instr{p}{q}$ to denote
the instrumentation of program $p$ where $q$ is the result
of instrumentation.
Function $\instrt{\cdot}$ is defined by structural induction
on a program.
We assume a similar instrumentation function for commands.
 \bda{lcl}
 \instrt{[]} & = & []
  \\
 \instrt{c:p} & = & \instrt{c} : \instrt{p}
 \\
 \\
  \instrt{y \assign b} & = & [y \assign b]
  \\
  \instrt{y \assign \SYNCMAKECHAN} & = & [y \assign \SYNCMAKECHAN]
  %% MS:
  %% a bit of cheating, no type info attached to synchronous channel
  \\
  \instrt{\GO\ p} & = & [\GO\ ([x_{tid} \assign [] \pp\ \instrt{p}])]
  \\
  \instrt{\SELECT\ [e_i \Rightarrow p_i]_{i\in \{1,\dots,n\}}}
     & =  & [x_{tid} \assign x_{tid} \pp\ [\pre{[\retrieve{e_1},\dots,\retrieve{e_n}]}],
         \\ & &\SELECT\ [\instrt{e_i \Rightarrow p_i}]_{i \in \{1,\dots,n\}}]
  \\
  \instrt{\SEND{x}{b} \Rightarrow p}
      & =  & \SEND{x}{[\tid,b]} \Rightarrow
                  (x_{tid} \assign x_{tid} \pp\ [\post{\snd{x}}]) \pp\ \instrt{p}
  \\                 
  \instrt{y \assign \RCV{x} \Rightarrow p}
   & = & y' \assign \RCV{x} \Rightarrow
          [x_{tid} \assign x_{tid} \pp\ [\post{\thread{\head(y')}{\rcv{x}}}],           
       \\ &&   \ \ \ \ \ \ \ \ \ \ \ \ \ \ \  y \assign \last(y')]   \pp\ \instrt{p}
  \eda
  
    \bda{c}
  \retrieve{\SEND{x}{b}}  =  \snd{x}
  \ \ \ \  \retrieve{y = \RCV{x}}  =  \rcv{x}
  \eda
\end{definition}

%% MS: omit, not much space left, example pretty much covers the below
%% For assignment and generation of channels,
%% instrumentation simply returns the original command.
%% In case of a \GO\ command, we introduce
%% a fresh thread local variable $x_{tid}$ to store
%% pre and post events in this thread.
%% Initially, $x_{tid}$ equals the empty list (trace).
%% In case of select, we first store all pre events
%% which are retrieved via another auxiliary function.
%% A select case performing a send is instrumented
%% by supplying the thread id of the sender.
%% In the body of the case, we store the respective post event.
%% For a receive case, we introduce a fresh variable $y'$.
%% This variables holds the senders thread id and the original value.
%% In the body, we store again the respective post event
%% and assign $y$ to the actually received value.

Run-time tracing proceeds as follows.
We simply run the instrumented program and
extract the local traces connected
to variables $x_{tid}$.
We assume that thread id numbers are created
during program execution and can be
enumerated by $1\dots n$
  for some $n>0$ where
  thread id number~$1$ belongs to the main thread.

\begin{definition}[Run-Time Tracing]
  Let $p$ and $q$ be programs such that $\instrt{p} = q$.
  We consider a specific instrumented program run
  where
  $\semP{\Config{()}{[\thread{1}{[x_1 \assign []] \pp\ q}]}}{T}{\Config{S}{\thread{1}{[]}:P}}$
  for some $S$, $T$ and $P$.
  Then, we refer to $T$ as $p$'s
  \emph{actual run-time trace}.
  We refer to the list $[\thread{1}{S(x_1)},\dots,\thread{n}{S(x_n)}]$
  as the \emph{local traces} obtained via the instrumentation
  of $p$.
\end{definition}
Command $x_1 \assign []$ is added to the instrumented
program to initialize the trace of the main thread.
Recall that main has thread id number~$1$.
This extra step is necessary because our instrumentation
only initializes local traces
of threads generated via \GO.
The final configuration $(S,\thread{1}{[]}:P)$ indicates
that the main thread has run to full completion.
This is a realistic assumption
as we assume that programs exhibit no obvious bug during execution.
There might still be some pending threads, in case $P$ differs
from the empty list.
%% In Go, such threads will be automatically terminated
%% once the main thread terminates.

%--------------------------------------------------------
%--------------------------------------------------------
\section{Trace Analysis}
\label{sec:trace-analysis}

We assume that the program has been instrumented
and after some program run we obtain
a list of local traces.
We show that the actual run-time trace
can be recovered  and we are able to point out alternative behaviors
that could have taken place.
Alternative behaviors are either due alternative schedules or different choices
among communication partners.

We consider the list of local traces 
$[\thread{1}{S(x_1)}, \dots, \thread{n}{S(x_n)}]$.
Their shape can be characterized as follows.

\begin{definition}[Local Traces]

  \bda{lcl}
  U,V & ::= & [] \mid \thread{i}{L} : U
  \\
  L & ::= & [] \mid \pre{as} : M
  \\
    as & ::= & [] \mid \snd{x} : as \mid \rcv{x} : as
    \\
    M & ::= & [] \mid \post{\snd{x}} : L
                  \mid \post{\thread{i}{\rcv{x}}} : L  
  \eda

  We refer to $U=[\thread{1}{L_1},\dots,\thread{n}{L_n}]$
  as a \emph{residual} list of local traces
  if for each $L_i$ either $L_i = []$ or $L_i=[\pre{\dots}]$.
\end{definition}

To recover the communications that took place
we check for matching pre and post events recorded
in the list of local traces.
For this purpose, we introduce a relation $\replay{U}{T}{V}$
to denote that `replaying' of $U$ leads to $V$
where communications $T$ took place.
Valid replays are defined via the following rules.

\begin{definition}[Replay $\replay{U}{T}{V}$]
\label{def:replay}
  \bda{c}

\rlabel{Sync} \
\myirule{L_1 = \pre{[\dots,\snd{x},\dots]} : \post{\snd{x}} : L_1'
        \\ L_2 = \pre{[\dots,\rcv{x}, \dots]} : \post{\thread{i_1}{\rcv{x}}} : L_2'}
        {\replay{\thread{i_1}{L_1} :
                 \thread{i_2}{L_2} : U}
          {[\sndEvt{i_1}{x}, \rcvEvt{i_2}{i_1}{x}]}
          {\thread{i_1}{L_1'} : \thread{i_2}{L_2'} : U}
        }
   \\
   \\
   \rlabel{Schedule} \
   \myirule{\mbox{$\pi$ permutation on $\{1,\dots,n\}$}}
           {\replay{[\thread{i_1}{L_1},\dots,\thread{i_n}{L_n}]}
                   {[]}
                   {[\thread{i_{\pi(1)}}{L_{\pi(1)}},\dots,\thread{i_{\pi(n)}}{L_{\pi(n)}}]}
           }
     \\
     \\
     \rlabel{Closure} \
     \myirule{\replay{U}{T}{U'} \ \ \replay{U'}{T'}{U''}}
             {\replay{U}{T \pp\ T'}{U''}}
  \eda
\end{definition}
Rule \rlabel{Sync} checks for matching
communication partners. In each trace, we must find complementary
pre events and the post events must match as well.
Recall that in the instrumentation the sender transmits its thread id
to the receiver.
Rule \rlabel{Schedule} shuffles the local traces
as rule \rlabel{Sync} only considers the two leading local traces.
Via rule \rlabel{Closure} we perform repeated replay steps.

We can state that the actual run-time trace
can be obtained via the replay relation $\replay{U}{T}{V}$
but further run-time traces are possible.
This is due to alternative schedules.

\begin{proposition}[Replay Yields Run-Time Traces]
  Let $p$ be a program and $q$ its instrumentation
  where for a specific program run we observe
  the actual behavior $T$ and the list $[\thread{1}{L_1},\dots,\thread{n}{L_n}]$
  of local traces.
  Let ${\cal T} = \{ T' \mid \replay{[\thread{1}{L_1},\dots,\thread{n}{L_n}]}{T'}{\thread{1}{[]}:U}
  \ \mbox{for some residual $U$} \}$.
  Then, we find that $T \in {\cal T}$
  and for each $T' \in {\cal T}$ we have that
  $\semP{\Config{()}{p}}{T'}{\Config{S}{\thread{1}{[]} : P}}$
  for some $S$ and $P$.
\end{proposition}

\begin{definition}[Alternative Schedules]
  We say $[\thread{1}{L_1},\dots,\thread{n}{L_n}]$ contains
  \emph{alternative schedules}
  iff the cardinality of the set $\{ T' \mid \replay{[\thread{1}{L_1},\dots,\thread{n}{L_n}]}{T'}{\thread{1}{[]}:U}
  \ \mbox{for some residual $U$} \}$
  is greater than one.
\end{definition}  

We can also check if even further run-time traces
might have been possible by testing for
alternative communications.

\begin{definition}[Alternative Communications]
We say $[\thread{1}{L_1},\dots,\thread{n}{L_n}]$ contains 
\emph{alternative matches} iff for some $i,j, x, L, L'$ we have that
(1) $L_i = \pre{[\dots,\snd{x},\dots]} : L$, 
(2) $L_j = \pre{[\dots,\rcv{x}, \dots]} : L'$, and
(3) if $L = \post{\snd{x}}:L''$ for some $L''$
then $L' \not = \post{\thread{j}{\rcv{x}}} : L'''$
for any $L'''$.

We say $U=[\thread{1}{L_1},\dots,\thread{n}{L_n}]$ contains 
\emph{alternative communications}
iff $U$ contains alternative matches or
there exists $T$ and $V$ such that
$\replay{U}{T}{V}$ and $V$ contains alternative matches.
\end{definition}

The alternative match condition states that a sender could synchronize
with a receiver (see (1) and (2))
but this synchronization did not take place (see (3)).
For an alternative match to result in an alternative
communication, the match must be along a possible run-time trace.

%--------------------------------------------------------
\subsection{Dependency Graph for Efficient Trace Analysis}
\label{sec:dep-graph}

%--------- figure -----------%
\begin{figure}[tp]

  \bda{l}
[x \assign\ \SYNCMAKECHAN,
  y \assign\ \SYNCMAKECHAN,
  \\
  \GO\ [z \assign\ \ploc{\RCV{y}}{6}],
  \GO\ [\ploc{\SEND{y}{1}}{4}, \ploc{\SEND{x}{1}}{5}],
  \GO\ [\ploc{\SEND{x}{1}}{3}],
  \\
  x \assign\ \ploc{\RCV{x}}{1},
  x \assign\ \ploc{\RCV{x}}{2}]
\eda

\vspace{-0.5cm}
\bda{cc}

\ba{l}
    [\thread{4}{[\pre{\ploc{\rcv{y}}{6}}, \post{\thread{3}{\ploc{\rcv{y}}{6}}}]},
      \\
      \ \thread{3}{[\pre{\ploc{\snd{y}}{4}}, \post{\ploc{\snd{y}}{4}},
          \pre{\ploc{\snd{x}}{5}}, \post{\ploc{\snd{x}}{5}}]},
      \\
      \ \thread{2}{[\pre{\ploc{\snd{x}}{3}}, \post{\ploc{\snd{x}}{3}}]},
      \\
      \ \thread{1}{[\pre{\ploc{\rcv{x}}{1}}, \post{\thread{2}{\ploc{\rcv{x}}{1}}},
                    \pre{\ploc{\rcv{x}}{2}}, \post{\thread{4}{\ploc{\rcv{x}}{3}}}]} ]
\ea
&
\ \ \ 
%%dot -Teps -o example1.eps example1.dot
\ba{c}
\includegraphics[scale=0.4]{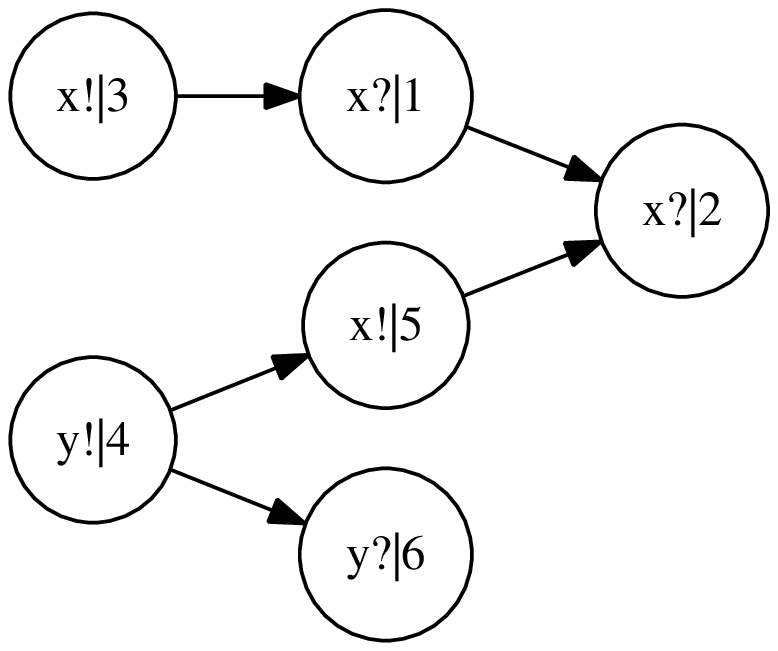}
\ea
\eda

  \caption{Dependency Graph among Events}
  \label{fig:dependency-graph}
\end{figure}

Instead of replaying traces to check for alternative schedules and
communications, we build a dependency graph
where the graph captures the partial order among events.
It is much more efficient to carry out the analysis on the graph
than replaying traces.
Figure~\ref{fig:dependency-graph} shows a simple example.

We find a program that makes use of two channels and four threads.
For reference, send/receive events are annotated (as subscript) with unique numbers.
We omit the details of instrumentation and assume that for a specific
program run we find the list of given traces on the left.
Pre events consist of singleton lists as there is no \SELECT.
Hence, we write $\pre{\ploc{\rcv{y}}{6}}$ as a shorthand for $\pre{[\ploc{\rcv{y}}{6}]}$.
Replay of the trace shows that the following
locations synchronize with each other:
$(4,6)$, $(3,1)$ and $(5,2)$.
This information as well as the order among events can
be captured by a dependency graph.
Nodes are obtained by a linear scan through the list of traces.
To derive edges, we require another scan for each element in a trace as we need to find pre/post pairs
belonging to matching synchronizations.
This results overall in $O(m*m)$
for the construction of the graph where $m$ is the number of elements found in each trace.
To avoid special treatment of 
dangling pre events (with not subsequent post event),
we assume that some dummy post events are added to the trace.

\begin{definition}[Construction of Dependency Graph]
\label{def:depend-graph}  
Each node corresponds to a send or a receive operation in the program text.
Edges are constructed by observing events recorded in the list of traces.
We draw a (directed) edge among nodes if either
\begin{itemize}
 \item the pre and post events
of one node precede the pre and post events of another node in the trace, or
\item the pre and post events belonging to both nodes can be synchronized.
      See rule \rlabel{Sync} in Definition~\ref{def:replay}.
      We assume that the edge starts from the node with the send operation.
\end{itemize}
\end{definition}
Applied to our example, this results in the graph on the right. See Figure~\ref{fig:dependency-graph}.
For example, $\snd{x}|3$ denotes a send communication over channel $x$ at program location~$3$.
As send precedes receive we find an edge from $\snd{x}|3$
to $\rcv{x}|1$. In general, there may be several initial nodes.
By construction, each node has at most one outgoing edge but may have multiple incoming edges.

The trace analysis 
can be carried out directly on the dependency graph.
To check if one event happens-before another event we seek for
a path from one event to the other.
This can be done via a depth-first search and takes time $O(v+e)$
where $v$ is the number of nodes and $e$ the number of edges.
Two events are concurrent if neither happens-before the other.
To check for alternative communications, we check for
matching nodes that are concurrent to each other.
By matching we mean that one of the nodes is a send and the other is a receive
over the same channel.
For our example, we find that
$\snd{x}|5$ and $\rcv{x}|1$ represents an alternative
communication as both nodes are matching
and concurrent to each other.

To derive (all) alternative schedules, we perform
a backward traversal of the graph.
Backward in the sense that we traverse the graph by moving from children to parent node.
We start with some final node (no outgoing edge).
Each node visited is marked.
We proceed to the parent if all children are marked.
Thus, we guarantee that the happens-before relation is respected.
For our example, suppose we visit first $\rcv{y}{6}$.
We cannot visit its parent $\snd{y}{4}$ until we have
visited $\rcv{x}{2}$ and $\snd{x}{5}$.
Via a (backward) breadth-first search we can `accumulate' all schedules.

%--------------------------------------------------------
%--------------------------------------------------------
\section{Comparison to Vector Clock Method}
\label{sec:comparison}

Via a simple adaptation of the Replay Definition~\ref{def:replay}
we can attach vector clocks to each send and receive event.
Hence, our tracing method strictly subsumes the vector clock method
as we are also able to trace events that could not commit.

\begin{definition}[Vector Clock]
\bda{lcll}
 cs & ::= & [] \mid n : cs  
\eda
\end{definition}
For convenience, we represent a vector clock as a list of clocks where the first position
belongs to thread 1 etc.
We write $cs[i]$ to retrieve the $i$-th component in $cs$.
We write $\incC{i}{cs}$ to denote the vector clock obtained from $cs$
where all elements are the same but at index $i$ the element is incremented by one.
We write $\maxC{cs_1}{cs_2}$ to denote the vector clock where we per-index take
the greater element.
We write $\clock{i}{cs}$ to denote thread $i$ with vector clock $cs$.
We write $\clock{\sndEvt{i}{x}}{cs}$ to denote a send over channel $x$ in thread $i$ with vector clock $cs$.
We write $\clock{\rcvEvt{i}{j}{x}}{cs}$ to denote a receive over channel $x$ in thread $i$ from thread $j$ with vector clock $cs$.

\begin{definition}[From Trace Replay to Vector Clocks]

  \bda{c}

\rlabel{Sync} \
\myirule{L_1 = \pre{[\dots,\snd{x},\dots]} : \post{\snd{x}} : L_1'
        \\ L_2 = \pre{[\dots,\rcv{x}, \dots]} : \post{\thread{i_1}{\rcv{x}}} : L_2'
        \\ cs = \maxC{\incC{i_1}{cs_1}}{\incC{i_2}{cs_2}}
        }
        {\replay{\thread{\clock{i_1}{cs_1}}{L_1} :
                 \thread{\clock{i_2}{cs_2}}{L_2} : U}
          {[\clock{\sndEvt{i_1}{x}}{cs}, \clock{\rcvEvt{i_2}{i_1}{x}}{cs}]}
          {\thread{\clock{i_1}{cs}}{L_1'} : \thread{\clock{i_2}{cs}}{L_2'} : U}
        }
  \eda        
\end{definition}
Like the construction of the dependency graph, the (re)construction of vector clocks takes time
$O(m*m)$ where $m$ is the number of elements found in each trace.

%% MS: said earlier
%% We can provide more information.
%% Alternative communication partners in case of partial commits.
%% See newsreader example in the introduction.

To check for an alternative communication,
the vector clock method seeks for matching events.
This incurs the same (quadratic in the size of the trace) cost as for our method.
However, the check that these two events are concurrent to each other
can be performed more efficiently via vector clocks.
Comparison of vector clocks takes time $O(n)$ where $n$ is the number of threads.
Recall that our graph-based method requires time $O(v+e)$
where $v$ is the number of nodes and $e$ the number of edges.
The number $n$ is smaller than $v+e$.

However, our dependency graph representation is more efficient in case of exploring alternative schedules.
In case of the vector clock method, we need to continuously compare vector clocks
whereas we only require a (backward) traversal of the graph.
We believe that the dependency graph has further advantages in case of
user interaction and visualization
as it is more intuitive to navigate through the graph.
This is something we intend to investigate in future work.

%--------------------------------------------------------
%--------------------------------------------------------
\section{Implementation}
\label{sec:implementation}

We have fully integrated the approach laid out in the earlier
sections into the Go programming language
and have built a prototype tool.
We give an overview of our implementation which can be found here~\cite{goscout}.
A detailed treatment of all of Go's message-passing features
can be found in the extended version of this paper.

%--------------------------------------------------------
\subsection{Library-Based Instrumentation and Tracing}

We use a pre-processor to carry out the instrumentation
as described in Section~\ref{sec:instrumentation}.
In our implementation, each thread maintains an entry in a lock-free hashmap
where each entry represents a thread (trace).
The hashmap is written to file either
at the end of the program or when a deadlock occurs.
We currently do not deal with the case that the program crashes
as we focus on the detection of potential bugs in programs
that do not show any abnormal behavior.
%% MS: todo
%% \ms{what if the program crashes?}

%--------------------------------------------------------
\subsection{Measurement of Run-Time Overhead Library-Based Tracing}
\label{sec:experiments}
%TODO
%
%set of benchmark examples
%
%what to measure?
%
%\begin{itemize}
%\item Overhead compared to original program.
%\item Overhead pre/post vs vector clocks.
%\item Size of traces? Analysis times?  
%\end{itemize}
%
%TODO: need best and worst cases, where performance of vector clocks similar
%to our approach, worst case, we're much better.
%
%
%TODO: Things to think about.
%Pre/pre vs vector clocks. We may encounter different program runs!
%Most likely, pre/post have smaller overhead, so we are closer to the actual program run!
%
%
%Further thoughts:
%We could extend vector clocks with pre events, would be then equally expressive
%as our approach. Could mention that somewhere.
%
%And even further thoughts:
%Could optimize vector clocks, don't transmit entire array, only the known/connected ones.
%\\

\begin{figure}
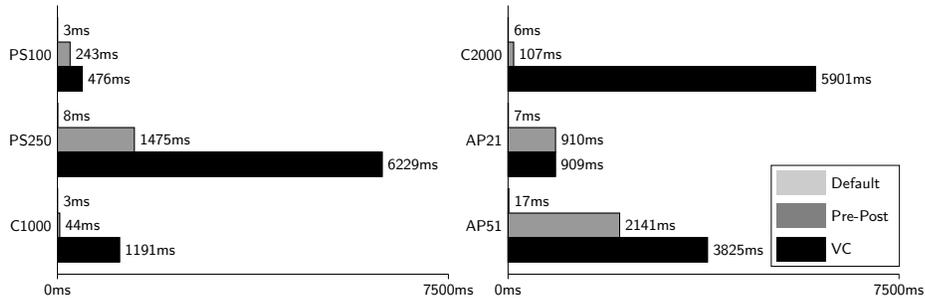

%% \begin{bchart} [max=7500,unit=ms]
%% 	\bcbar[color=black!20]{3}
%% 	\bcbar[label=Primesieve100, color=black!40]{243}
%% 	\bcbar[color=black]{476}
%% 	\smallskip
%% 	\bcbar[color=black!20]{8}
%% 	\bcbar[label=Primesieve250, color=black!40]{1475}
%% 	\bcbar[color=black]{6229}
%% 	\smallskip
%% 	\bcbar[color=black!20]{3}
%% 	\bcbar[label=Collector1000, color=black!40]{44}
%% 	\bcbar[color=black]{1191}
%% 	\smallskip
%% 	\bcbar[color=black!20]{6}
%% 	\bcbar[label=Collector2000, color=black!40]{107}
%% 	\bcbar[color=black]{5901}
%% 	\smallskip
%% 	\bcbar[color=black!20]{7}
%%     \bcbar[label=Add-Pipe21, color=black!40]{910}
%%     \bcbar[color=black]{909}
%%     \smallskip
%% 	\bcbar[color=black!20]{17}
%% 	\bcbar[label=Add-Pipe51, color=black!40]{2141}
%% 	\bcbar[color=black]{3825}
%% 
%% 	\bclegend{5pt}{black!20/Default,black!50/Pre-Post,black/VC}
%% \end{bchart}
  \scalebox{0.65}{
    \begin{tabular}{ll}
%%      \begin{minipage}[b]{.47\linewidth}
\begin{bchart} [max=7500,unit=ms]
	\bcbar[color=black!20]{3}
	\bcbar[label=PS100, color=black!40]{243}
	\bcbar[color=black]{476}
	\smallskip
	\bcbar[color=black!20]{8}
	\bcbar[label=PS250, color=black!40]{1475}
	\bcbar[color=black]{6229}
	\smallskip
	\bcbar[color=black!20]{3}
	\bcbar[label=C1000, color=black!40]{44}
	\bcbar[color=black]{1191}
\end{bchart}
%%      \end{minipage}
      &  
%%      \begin{minipage}[b]{.47\linewidth}
 \hspace{-0.8cm}
\begin{bchart} [max=7500,unit=ms]
	\bcbar[color=black!20]{6}
	\bcbar[label=C2000, color=black!40]{107}
	\bcbar[color=black]{5901}
	\smallskip
	\bcbar[color=black!20]{7}
    \bcbar[label=AP21, color=black!40]{910}
    \bcbar[color=black]{909}
    \smallskip
	\bcbar[color=black!20]{17}
	\bcbar[label=AP51, color=black!40]{2141}
	\bcbar[color=black]{3825}

	\bclegend{5pt}{black!20/Default,black!50/Pre-Post,black/VC}
\end{bchart}
%%      \end{minipage}      
\end{tabular}
}

  \caption{Performance overhead using Pre/Post vs Vector clocks(VC) in ms.}
\label{fig:experiments}
\end{figure}
%\begin{tikzpicture}
%\begin{axis}[title  = Performance overhead using Pre/Post vs Vector clocks(VC) in ms,
%xbar,
%y axis line style = { opacity = 0 },
%axis x line       = none,
%tickwidth         = 0pt,
%enlarge y limits  = -0.5,
%%enlarge x limits  = 0.02,
%symbolic y coords = {Collector, Primesieve250, Primesieve, Add-Pipe51, Add-Pipe21, Add-Pipe11},
%nodes near coords,
%]
%\addplot coordinates { (4,Add-Pipe11)  (7,Add-Pipe21) (17,Add-Pipe51)     (23,Primesieve)   (4,Primesieve250)
%	(3,Collector)    };
%\addplot coordinates { (452,Add-Pipe11)   (910,Add-Pipe21) (2141,Add-Pipe51)   (243,Primesieve)   (1475,Primesieve250)
%	(44,Collector)     };
%\addplot coordinates { (398,Add-Pipe11)    (909,Add-Pipe21)  (3825,Add-Pipe51)   (476,Primesieve)  (6229,Primesieve250)
%	(1191,Collector)     };
%\legend{Default, Pre/Post, VC}
%\end{axis}
%\end{tikzpicture}

We measure the run-time overhead of our method against the vector clock method.
Both methods are implemented as libraries assuming
no access to the Go run-time system.
For experimentation we use three programs where
each program exercises some of the factors that have an impact on tracing.
For example, dynamic versus static number of threads and channels.
Low versus high amount of communication among threads.

%% MS: omit, need to be careful, not sure exactly what/how things have been optimized
%% Vector clock optimizations provide for constant factor improvement
%% relying on the assumption that the number of threads and channels is statically known.
%% As our experiments show, our approach scales well for examples
%% with a dynamic number of threads and channels whereas
%% the vector clock methods becomes impractical.

The \textbf{Add-Pipe} (\textbf{AP}) example uses $n$ threads where the first $n-1$ threads receive on an input channel,
add one to the received value and then send the new value on their output channel to
the next thread.
The first thread sends the initial value and receives the result from the last thread.

In the \textbf{Primesieve} (\textbf{PS}) example, the communication among threads is similar
to the \textbf{Add-Pipe} example. The difference is that
threads and channels are dynamically generated to calculate the first $n$ prime numbers.
For each found prime number a `filter' thread is created.
Each thread has an input channel
to receive new possible prime numbers $v$ and an output channel
to report each number for which $v \mod {\mathit prime} \neq 0$ where ${\mathit prime}$
is the prime number associated with this filter thread.
The filter threads are run in a chain where the first thread stores the prime number 2.
%% The amount of threads is $n+1$.

The \textbf{Collector} (\textbf{C}) example creates $n$ threads that
produce a number which is then sent to the main thread for collection.
This example has much fewer communications compared to the other examples
but uses a high number of threads.

Figure~\ref{fig:experiments} summarizes our results.
Results are carried out on some commodity hardware
(Intel i7-6600U with 12 GB RAM, a SSD and Go 1.8.3 running on Windows 10 was used for the tests).
Our results show that a library-based implementation of the vector clock method
does not scale well for examples with a dynamic number of threads and/or a high amount communication among threads.
See examples \textbf{Primesieve} and \textbf{Add-Pipe}.
None of the vector clock optimizations~\cite{DBLP:journals/dc/GargSM07}
apply here because of the dynamic number of threads and channels.
Our method performs much better. This is no surprise as we require less (tracing) data and no extra
communication links. We believe that the overhead can still be further reduced
as access to the thread id in Go is currently rather cumbersome and expensive.

\section{Conclusion}

%% MS: omit, no space
%% \ms{highlight, this can be quite useful to identify bugs, e.g. send might happen after close. See examples.
%% Similar for select, can show alternative cases might exist (not a bug directly).}

One of the challenges of run-time verification in the concurrent
setting is to establish a partial order among recorded events.
Thus, we can identify potential bugs due to bad schedules
that are possible but did not take place in some specific program run.
Vector clocks are the predominant method to achieve this task.
For example, see
%% MS: omit, not cited by other data-race works, guess main focus not on data races?
%% work by Sen, Rosu and Agha~\cite{Sen:2003:RSA:949952.940116}
%% in the Java shared memory setting,
work by Vo~\cite{Vo:2011:SFD:2231450} in the MPI setting
and work by Tasharofi~\cite{Samira:Tasharofi:2013} in the actor setting.
There are several works that employ vector clocks in the shared memory setting
For example, see Pozniansky's and Schuster's work~\cite{DBLP:journals/concurrency/PoznianskyS07} on data race detection.
Some follow-up work by Flanagan and Freund~\cite{Flanagan:2009:FEP:1542476.1542490}
employs some optimizations to reduce the tracing overhead
by recording only a single clock instead of the entire vector.
We leave to future work to investigate whether such
optimizations are applicable in the message-passing setting
and how they compare to existing optimizations such as~\cite{DBLP:journals/dc/GargSM07}.

We have introduced a novel tracing method that has much less overhead
compared to the vector clock method. Our method can deal
with all of Go's message-passing language features
and can be implemented efficiently as a library.
We have built a prototype that can automatically identify
alternative schedules and communications.
In future work we plan to conduct some case studies
and integrate heuristics for specific
scenarios, e.g.~reporting a send operation on a closed channel etc.

%--------------------------------------------------------
%--------------------------------------------------------
\section*{Acknowledgments}

We thank some HVC'17 reviewers for their constructive feedback on an earlier version
of this paper.

\bibliography{main}

%% submit without appendix
%% \end{document}

\pagebreak

\appendix

%--------------------------------------------------------
%--------------------------------------------------------
\section{Further Go Message-Passing Features}

%--------------------------------------------------------
\subsection{Overview}

%--------- figure -----------%
\begin{figure}[tp]

\begin{tabular}{ll}

  \begin{minipage}[b]{.47\linewidth}

\begin{lstlisting}    
func A(x chan int) {
  x <- 1       // A1
}

func bufferedChan() {
  x := make(chan int,1)
  go A(x)
  x <- 1      // A2
  <-x
}

func closedChan() {
  x := make(chan int)
  go A(x)
  go B(x)
  close(x)
}
\end{lstlisting}
  \end{minipage}
  
  \begin{minipage}[b]{.47\linewidth}
    \begin{lstlisting}
func B(x chan int) {
  <-x
}
      
func selDefault() {
  x := make(chan int)
  go A(x)

  select {
  case <-x:      // A3
    fmt.Println("received from x")
  default:
    fmt.Println("default")
  }
}
\end{lstlisting}
    
  \end{minipage}

\end{tabular}  
  
  \caption{Further Go Features}
  \label{fig:further-go}
\end{figure}  

%% For space reasons, our formal description only covers synchronous channels
%% and selective communications.
Besides selective synchronous message-passing,
Go supports some further message passing features
that can be easily dealt with by our approach and are fully supported
by our implementation.
Figure~\ref{fig:further-go} shows such examples where
we put the program text in two columns.

\paragraph{Buffered Channels}

 %% MS: omit, one single fig instead
 %% \begin{lstlisting}[float={tp},captionpos={b}, caption={Buffered Channel}, label={lst:buffer}]
 %% func A(x chan int) {
 %%   x <- 1                    // A1
 %% }
 %% func main() {
 %%   x := make(chan int,1)
 %%   go A(x)
 %%   x <- 1                    // A2
 %%   <-x
 %% }
 %% \end{lstlisting}
 
 Go also supports buffered channels
 where send is asynchronous assuming sufficient buffer space exists.
 See function \texttt{buffered} in Figure~\ref{fig:further-go}. %%Listing~\ref{lst:buffer}.
 Depending on the program run, our analysis reports that either A1 or A2 are
 alternative matches for the receive operation.
 
 In terms of the instrumentation and tracing, we treat each asynchronous send
 as if the send is executed in its own thread.
 %% Of course, the actual asynchronous send operation may possibly block.
 This may lead to some slight inaccuracies.
 Consider the following variant.
 \begin{lstlisting}
 func buffered2() {
   x := make(chan int,1)
   x <- 1                    // B1         
   go A(x)                   // B2
   <-x                       // B3
 }
 \end{lstlisting}
 Our analysis  reports that B2 and B3 form an alternative match.
 However, in the Go semantics, buffered messages are queued.
 Hence, for \emph{every} program run the only possibility
 is that B1 synchronizes with B3. B3 never takes place!
 As our main objective is bug finding, we argue that
 this loss of accuracy is justifiable.
 How to eliminate such false positives is subject of future work.

\paragraph{Select with default/timeout}

%% MS: omit, one single fig instead
%% \begin{lstlisting}[float={tp},captionpos={b}, caption={Select with Default}, label={lst:selectCode}]
%% func A(x chan int) {
%%   x <- 1
%% }
%% func main() {
%%   x := make(chan int)
%%   go A(x)
%% 
%%   select {
%%   case <-x:
%%     fmt.Println("received from x")
%%   default:
%%     fmt.Println("default")
%%   }
%% }
%% \end{lstlisting}

Another feature in Go is to include a default/timeout case to \SELECT.
See \texttt{selDefault} in Figure~\ref{fig:further-go}.
The purpose is to avoid (indefinite) blocking if none of the other cases are available.
For the user it is useful to find out if other alternatives are available
in case the default case is selected.
%%Consider the program in Listing~\ref{lst:selectCode}.
The default case applies for most program runs.
Our analysis reports that A1 and A3 are an alternative match.
%% MS: omit, not clear if this is the actual output
%% %
%% \begin{quote}
%%      Alternatives for main,[(1(0),?,P,select2.go:8),(default,\_,\_,select2.go:10)]
%%      {\color{red}A,[(1(0),!,P,select2.go:2)]}
%% \end{quote}     
%% %

To deal with default/timeout we introduce
a new post event $\post{\select}$.
To carry out the analysis in terms of the dependency graph,
each subtrace $\dots,\pre{[\dots,\select,\dots]},\post{\select},\dots$
creates a new node.
Construction of edges remains unchanged.

\paragraph{Closing of Channels}

%% MS: omit, single fig instead
%% \begin{lstlisting}[float={tp},captionpos={b}, caption={Closing of channels example}, label={lst:closingCode}]
%% func A(x chan int) {
%%   x <- 1
%% }
%% func B(x chan int) {
%%   <-x
%% }
%% func main() {
%%   x := make(chan int)
%%   go A(x)
%%   go B(x)
%%   close(x)
%% }
%% \end{lstlisting}

Another feature in Go is the ability to close a channel.
See \texttt{closedChan} in Figure~\ref{fig:further-go}.
Once a channel is closed, each send on a closed channel
leads to failure (the program crashes).
On the other hand, each receive on a closed channel is always successful,
as we receive a dummy value.
%%Consider the program in Listing~\ref{lst:closingCode}.
A run of is successful if the close operation of the main thread
happens after the send in thread \texttt{A}.
As the close and send operations happen concurrently,
our analysis reports that the send A1 may take place after close.

For instrumentation/tracing, we introduce event $\close{x}$.
It is easy to identify a receive on a closed channel,
as we receive a dummy thread id.
So, for each subtrace
$[\dots,\pre{[\dots,\rcv{x},\dots]},\post{\thread{i}{\rcv{x}}},\dots]$
where $i$ is a dummy value we draw an edge
from $\close{x}$ to $\rcv{x}$.

%% MS: omit, analysis report?
%% \begin{figure}
%% 	\footnotesize
%% 	Actions parallel or after main,[(1(0),\#,C,close.go:11)]\\
%% 	{\color{OliveGreen}B,[(1(0),?,P,close.go:5)]}\\
%% 	{\color{red}A,[(1(0),!,P,close.go:2)]}
%% 	\caption{Analysis result for example \ref{closingCode}. \texttt{main} is the thread that triggers the event, \texttt{1} is the channel ID, (0) the buffer size where 0 means synchronous and !/?/\# are the symbols for send/receive/close. The parallel send is an error which could have crashed the program if another schedule was used.}
%% \end{figure}

Here are the details of how to include buffered channels, select and closing of channels.

%--------------------------------------------------------
\subsection{Buffered Channels}

Consider the following Go program.
\begin{lstlisting}
  x := make(chan, 2)
  x <- 1    // E1
  x <- 1    // E2
  <- x      // E3
  <- x      // E4
\end{lstlisting}
We create a buffer of size 2. The two send operations
will then be carried out asynchronously and
the subsequent receive operations will pick up the buffered values.
We need to take special care of buffered send operations.
If we would treat them like synchronous send operations,
their respective pre and post events would be recorded
in the same trace as the pre and post events of the receive operations.
This would have the consequence that our trace analysis
does not find out that events E1 and E2 happen before E3 and E4.

Our solution to this issue is to treat each send operation on a buffered channel
as if the send operation is carried out in its own
thread. Thus, our trace analysis is able to detect that E1 and E2 take place before E3 and E4.
This is achieved by marking each send on a buffered channel in the instrumentation.
After tracing, pre and post events will then be moved to their own trace.
From the viewpoint of our trace analysis, a buffered channel then appears
as having infinite buffer space. Of course, when running the program
a send operation may still block if all buffer space is occupied.

Here are the details of the necessary adjustments to our method.
During instrumentation/tracing, we simply record if a buffered send operation took place.
The only affected case in the instrumentation of commands (Definition~\ref{def:instrumentation})
is $\SEND{x}{b} \Rightarrow p$.
We assume a predicate $\isBuffered{\cdot}$ to check if a channel is buffered or not.
In terms of the actual implementation this is straightforward to implement.
We write $\postAsync{x}{n}$ to indicate a buffered send operation via $x$
where $n$ is a fresh thread id.
We create fresh thread id numbers via $\newTID$. 

\begin{definition}[Instrumentation of Buffered Channels]
Let $x$ be a buffered channel.
  \bda{lcl}
  \instrt{\SEND{x}{b} \Rightarrow p}
  &&
  \\
  \ \ \mid \isBuffered{x}    & =  & \SEND{x}{[n,b]} \Rightarrow
  (x_{tid} \assign x_{tid} \pp\ [\postAsync{\snd{x}{n}}]) \pp\ \instrt{p}
  \\ & & \mbox{where} \ \ n = \newTID
  \\
  \ \ \mid \mbox{otherwise} & = & \SEND{x}{[\tid,b]} \Rightarrow
                  (x_{tid} \assign x_{tid} \pp\ [\post{\snd{x}}]) \pp\ \instrt{p}
 \eda  
\end{definition}

The treatment of buffered channels has no overhead on the instrumentation and tracing.
However, we require a post-processing phase where
marked events will be then moved to their own trace.
This can be achieved via a linear scan through each trace.
Hence, requires time complexity $O(k)$ where $k$
is the overall size of all (initially recorded) traces.
For the sake of completeness, we give below
a declarative description of post-processing
in terms of relation $\postProc{U}{V}$.

\begin{definition}[Post-Processing for Buffered Channels $\postProc{U}{V}$]
  \bda{c}
  \rlabel{MovePostB} \
  \myirule{ L = \pre{as} : \postAsync{\snd{x}}{n} : L'}
          {\postProc{\thread{i}{L} : U}
                    {\thread{i}{L'} : \thread{n}{[\pre{as}, \postAsync{\snd{x}}{n}]} : U}}
  \\
  \\
  \rlabel{Shift} \
  \myirule{L = \pre{as} : \post{a} : L'
          \\ (a = \snd{x} \ \vee\ a = \thread{j}{\rcv{x}})
          \\ \postProc{\thread{i}{L'}:U}{\thread{i}{L''}:U'} }
          {\postProc{\thread{i}{L} : U}{\thread{i}{\pre{as} : \post{a} : L''} : U'}}
  \\
  \\        
   \rlabel{Schedule} \
   \myirule{\mbox{$\pi$ permutation on $\{1,\dots,n\}$}}
           {\postProc{[\thread{i_1}{L_1},\dots,\thread{i_n}{L_n}]}
                   {[\thread{i_{\pi(1)}}{L_{\pi(1)}},\dots,\thread{i_{\pi(n)}}{L_{\pi(n)}}]}
           }
     \\
     \\
     \rlabel{Closure} \
     \myirule{\postProc{U}{U'} \ \ \postProc{U'}{U''}}
             {\postProc{U}{U''}}          
  \eda
\end{definition}

Subsequent analysis steps will be carried out on the list of traces obtained via post-processing.

There is some space for improvement.
Consider the following program text.
\begin{lstlisting}
  func A(x chan int) {
  x <- 1       // A1
}
func buffered2() {
  x := make(chan int,1)
  x <- 1                    // B1         
  go A(x)                   // B2
  <-x                       // B3
}
\end{lstlisting}
Our analysis (for some program run) reports that B2 and B3 is an alternative match.
However, in the Go semantics, buffered messages are queued.
Hence, for \emph{every} program run the only possibility
is that B1 synchronizes with B3. B3 never takes place.
As our main objective is bug finding, we can live with this inaccuracy.
We will investigate in future work how to eliminate this false positive.

%--------------------------------------------------------
%--------------------------------------------------------
\section{Select with default/timeout}

In terms of the instrumentation/tracing, we introduce a new special post event
$\post{\select}$.
For the trace analysis (Definition~\ref{def:replay}), we require a new rule.

\bda{c}
\rlabel{Default/Timeout} \
\replay{\thread{i}{\pre{[\dots]}: \post{\select} : L} : U}
       {[\thread{i}{\select}]}
       {\thread{i}{L} : U}
\eda
This guarantees
that in case default or timeout is chosen, select acts as if asynchronous.

The dependency graph construction
easily takes care of this new feature.
For each default/timeout case we introduce a node.
Construction of edges remains unchanged.

%--------------------------------------------------------
%--------------------------------------------------------
\section{Closing of Channels}
\label{sec:closing-channels}

For instrumentation/tracing of the $\close{x}$ operation
on channel $x$, we introduce a special pre and post event.
Our trace analysis keeps track of closed channels.
As a receive on a closed channel yields some dummy values,
it is easy to distinguish this case from the regular \rlabel{Sync}.
Here are the necessary adjustments to our replay relation
from Definition~\ref{def:replay}.

\bda{lcl}
 C & ::= & [] \mid \thread{i}{\close{x}} : C
\eda

\bda{c}
\rlabel{Close} \
\replay{\sep{\thread{i}{\pre{\close{x}} : \post{\close{x}} : L} : U}{C}}
       {[]}
       {\sep{\thread{i}{L} : U}{\thread{i}{\close{x}} : C}}
\\
\\
\rlabel{RcvClosed} \
\myirule{Q = \thread{j}{\close{x}} : Q'
        }
        {
          \replay{\sep{\thread{i}{\pre{[\dots,\rcv{x},\dots]} : \post{\thread{j'}{\rcv{x}}} : L} : U}{Q}}
                 {[\thread{j}{\close{x}}, \rcvEvt{i}{j}{x}]}
                 {\sep{\thread{i}{L} : U}{Q}}
        }
\eda

For the construction of the  dependency graph,
we create a node for each close statement.
For each receive on a closed channel $x$ at program location $l$, we draw an edge from $\close{x}$
to $\rcv{x}|l$.

%--------------------------------------------------------
%--------------------------------------------------------
\section{Codes used for the Experimental results}

\subsection*{Add-Pipe}
\begin{lstlisting}[frame=single]
func add1(in chan int) chan int {
  out := make(chan int)
  go func() {
    for {
      n := <-in
      out <- n + 1
    }
  }()
  return out
}

func main() {
  in := make(chan int)
  c1 := add1(in)
  for i := 0; i < 19; i++ {
    c1 = add1(c1)
  }
  for n := 1; n < 1000; n++ {
    in <- n
    <-c1
  }
}
\end{lstlisting}

\subsection*{Primesieve}
\begin{lstlisting}[frame=single]
func generate(ch chan int) {
  for i := 2; ; i++ {
    ch <- i
  }
}

func filter(in chan int, out chan int, prime int) {
  for {
    tmp := <-in
    if tmp%prime != 0 {
      out <- tmp
    }
  }
}

func main() {
  ch := make(chan int)
  go generate(ch)
  for i := 0; i < 100; i++ {
    prime := <-ch
    ch1 := make(chan int)
    go filter(ch, ch1, prime)
    ch = ch1
  }
}
\end{lstlisting}

\subsection*{Collector}
\begin{lstlisting}[frame=single]
func collect(x chan int, v int) {
  x <- v
}
func main() {
  x := make(chan int)
  for i := 0; i < 1000; i++ {
    go collect(x, i)
  }

  for i := 0; i < 1000; i++ {
    <-x
  }
}
\end{lstlisting}

\end{document}